\documentclass[epj,nopacs]{svjour}
\usepackage{graphics}
\usepackage{xcolor}

\begin{document}

\title{\boldmath Study of the process
$e^+e^- \to \eta\pi^0\gamma$ in the energy range $\sqrt{s} = \mbox{1.05--2.00}$ GeV with the SND detector}
\authorrunning{M.~N.~Achasov et al.}
\titlerunning{Process
$e^+e^- \to \eta\pi^0\gamma$ in the energy range $\sqrt{s} = \mbox{1.05--2.00}$ GeV}

\author{{\large The SND Collaboration}\\ \\
M.~N.~Achasov\inst{1,2} \and
A.~Yu.~Barnyakov\inst{1,2} \and
A.~A.~Baykov\inst{1,2} \and
K.~I.~Beloborodov\inst{1,2} \and
A.~V.~Berdyugin\inst{1,2} \and
D.~E.~Berkaev\inst{1,2} \and
A.~G.~Bogdanchikov\inst{1}\and
A.~A.~Botov\inst{1}\and
T.~V.~Dimova\inst{1,2} \and
V.~P.~Druzhinin\inst{1,2} \and
V.~B.~Golubev\inst{1} \and
A.~N.~Kirpotin\inst{1} \and
L.~V.~Kardapoltsev\inst{1,2} \and
A.~S.~Kasaev\inst{1} \and
A.~G.~Kharlamov\inst{1,2} \and
I.~A.~Koop\inst{1,2} \and
A.~A.~Korol\inst{1,2} \and
D.~P.~Kovrizhin\inst{1} \and
A.~S.~Kupich\inst{1}\and
K.~A.~Martin\inst{1}\and
N.~A.~Melnikova\inst{1}\and
N.~Yu.~Muchnoy\inst{1,2} \and
A.~E.~Obrazovsky\inst{1}\and
E.~V.~Pakhtusova\inst{1}\and
K.~V.~Pugachev\inst{1,2} \and
D.~V.~Rabusov\inst{1}\and
Yu.~A.~Rogovsky\inst{1,2} \and
Y.~S.~Savchenko\inst{1,2} \and
A.~I.~Senchenko\inst{1,2} \and
S.~I.~Serednyakov\inst{1,2} \and
D.~N.~Shatilov\inst{1}\and
Yu.~M.~Shatunov\inst{1,2} \and 
D.~A.~Shtol\inst{1} \and
D.~B.~Shwartz\inst{1,2} \and
Z.~K.~Silagadze\inst{1,2} \and
I.~K.~Surin\inst{1} \and
M.~V.~Timoshenko\inst{1} \and
Yu.~V.~Usov\inst{1} \and
V.~N.~Zhabin\inst{1} \and
V.~V.~Zhulanov\inst{1,2}}

\institute{Budker Institute of Nuclear Physics, SB RAS, Novosibirsk, 630090, Russia \and 
Novosibirsk State University, Novosibirsk, 630090, Russia 
}
\date{}
\abstract{
The process $e^+e^-\to\eta\pi^0\gamma$ is studied in the center-of-mass energy
range 1.05–-2.00 GeV using data with an integrated luminosity of 94.5 pb$^{-1}$
collected by the SND detector at the VEPP-2000 $e^+e^−$ collider. 
The $e^+e^-\to\eta\pi^0\gamma$ cross section is  measured for the first time.
It is shown that the dominant mechanism of this reaction is the transition
through the $\omega\eta$ intermediate state. The measured cross section of the
subprocess $e^+e^-\to\omega\eta\to\eta\pi^0\gamma$ is consistent with previous
measurements in the $e^+e^-\to\pi^+\pi^-\pi^0\eta$ mode. It is found, with a
significance of 5.6$\sigma$, that the process $e^+e^-\to\eta\pi^0\gamma$ is
not completely described by hadronic vector-pseudoscalar intermediate states.
The cross section of this missing contribution, which can originate from 
radiation processes, e. g. $e^+e^-\to a_{0}(1450)\gamma$, is measured. It is
found to be 15--20 pb in the wide energy range from 1.3 to 1.9 GeV.}

\maketitle

\section{Introduction}
This work is devoted to study of the process
\begin{equation}
e^+e^- \to \eta\pi^0\gamma
\label{epg-eq}
\end{equation}
in the center-of-mass energy range $\sqrt{s} = 1.05$--2.00~GeV at the 
experiment with the SND detector at the VEPP-2000 $e^+e^-$ collider.
Previously, this process was studied near the $\phi$-meson resonance
by the SND at VEPP-2M~\cite{epg-phi-snd}, CMD-2~\cite{epg-phi-cmd2}
and KLOE~\cite{epg-phi-kloe}. The dominant intermediate mechanism in this 
energy region is the decay $\phi\to a_{0}(980)\gamma$.
Below ($\sqrt{s}=0.920$--1.004~GeV) and above ($\sqrt{s} = 1.03$--1.38~GeV)
the $\phi$-meson resonance the process~(\ref{epg-eq}) was studied by CMD-2 in
Ref.~\cite{epg-mhad-cmd2}, where a 90\% confidence-level upper limit of about
0.1 nb was set on the cross section. At higher energy, there is only the BESIII
measurement of the $J/\psi\to\eta\pi^0\gamma$ decay~\cite{epg-jpsi-bes}. 
The $\eta\pi^0$ mass spectrum in this decay is well described by an uniform
phase-space distribution. No significant signal from the $J/\psi$ decays
to $a_{0}(980)\gamma$ and $a_{2}(1320)\gamma$ was observed.

The dominant contribution to the $e^+e^-\to\eta\pi^0\gamma$ cross section
in the energy region under study comes from the process 
$e^+e^-\to\omega\eta$ with the decay $\omega\to\pi^0\gamma$. This process was
measured in the BABAR~\cite{ometa-babar}, CMD-3~\cite{eta3pi-cmd} and
SND~\cite{eta3pi-snd} experiments with the decay mode 
$\omega\to \pi^+\pi^-\pi^0$. In these works, the measurement was performed 
neglecting the interference between the $\omega\eta$ and other intermediate 
mechanisms ($a_{0}(980)\rho$, $\rho(1450)\pi$ and $\phi\eta$)
contributing to the  $e^+e^-\to\pi^+\pi^-\pi^0\eta$ reaction.
The measurements obtained under this assumption need additional verification,
especially in the region $\sqrt{s}=1.85$--2.00~GeV, where 
the $e^+e^-\to\omega\eta$ cross section is almost zero ($<50$ pb)
compared to the significant ($\sim$ 2 nb) contribution from other mechanisms.

In this work, the most interesting is the search for radiation decays of 
excited vector mesons of the $\rho$, $\omega$ and $\phi$ families to 
$a_{0}(980)\gamma$, $a_{2}(1320)\gamma$, and $a_{0}(1450)\gamma$. The 
measurement of these decays is important for understanding the quark structure
of excited vector mesons. In particular, there are indications 
that the excited states of the $\rho$ and $\omega$ mesons may contain
an admixture of a vector hybrid state~\cite{kalashnikova-hybrid}. 
The widths of the radiation decays are sensitive to the hybrid 
admixture~\cite{kalashnikova}.

\section{Detector and experiment}

SND is a general-purpose non-magnetic detector~\cite{SND_desc} collecting data
at the VEPP-2000 $e^+e^-$ collider~\cite{vepp2k}. Its main part is a 
three-layer spherical electromagnetic calorimeter consisting of 1630 NaI(Tl) 
crystals. The calorimeter covers a solid angle of 95\% of 4$\pi$. The energy
resolution of the calorimeter for photons is
$\sigma_{E}/E=4.2\%/\sqrt[4]{E(GeV)}$. The angular resolution is about
$1.5^\circ$. Directions of charged particles are measured using a nine-layer 
drift chamber and one-layer proportional chamber in a common gas volume. The
solid angle of the tracking system is 94\% of 4$\pi$. A system of threshold
aerogel Cherenkov counters located between the tracking system and the 
calorimeter is used for charged kaon identification. Outside the calorimeter,
a muon detector consisting of proportional tubes and scintillation counters
is placed.

Monte-Carlo (MC) simulation of the signal and the background processes takes
into account
radiative corrections~\cite{FadinRad}. The angular distribution of hard photon
emitted from the initial state is generated according to 
Ref.~\cite{BoneMartine}. Interactions of the particles produced in 
$e^+e^-$ annihilation with the detector materials are simulated using
the GEANT4 software~\cite{geant}. The simulation takes into account variation
of experimental conditions during data taking, in particular dead detector 
channels and beam-induced background. To take into account the effect of 
superimposing the beam background on the $e^+e^-$ annihilation events,
simulation uses special background events recorded during data taking with a 
random trigger. These events are superimposed on simulated events, 
leading to the appearance of additional tracks and photons in events.

The analysis presented in this work is based on data with an
integrated luminosity of 94.5 pb$^{-1}$ recorded in 2010, 2011,
2012 and 2017. These data were collected at 101 energy points in the
energy region $\sqrt{s} = 1.05$--2.00~GeV. Since the cross section of
the process under study is small and relatively slowly changes with energy,
the data are combined into 13 energy intervals shown in Table~\ref{nonomegatab}.

In this work, the process $e^+e^-\to\eta\pi^0\gamma$ is studied in the 
five-photon final state. Therefore, it is viable to use the process 
$e^+e^-\to\gamma\gamma$ for normalization. 
As a result of the
normalization a part of systematic uncertainties associated
with the hardware event selection and the beam background are canceled out.
Accuracy of the luminosity measurement using the 
process $e^+e^-\to\gamma\gamma$ is estimated to be 2\%.

\section{Event selection}
Selection of $e^+e^-\to\eta\pi^0\gamma\to5\gamma$ events is performed in two 
stages. At the first stage, we select events with exactly 5 photons with
energies above 20 MeV and no charged tracks. The latter condition is ensured by
requiring that the number of hits in the drift chamber is less than four.
The conditions on the total energy deposition in the calorimeter ($E_{\rm EMC}$)
and the  total event momentum ($P_{\rm EMC}$) calculated using energy 
depositions in calorimeter crystals are imposed:
\begin{equation}
E_{\rm EMC} / \sqrt{s} > 0.6, \quad P_{\rm EMC} / \sqrt{s} < 0.3.
\end{equation}
To suppress cosmic-ray background, absence of a signal from the muon system is
required.

Most events selected at this stage come from the process
\begin{equation}
\label{bkglist1}
e^+e^-\to\omega\pi^0\to\pi^0\pi^0\gamma.
\end{equation}
This background process was studied by SND in 
Ref.~\cite{SND_ompi}.
A noticeable contribution to the background 
comes also from the QED  processes
\begin{equation}
\label{bkglist2}
e^+e^- \to 3\gamma,\, 4\gamma,\, 5\gamma.
\end{equation}
We also study background from the following reactions with multiphoton
final states:
\begin{eqnarray}
&&e^+e^- \to \eta\gamma, \quad e^+e^- \to \pi^0\gamma,\nonumber\\
&&e^+e^- \to \omega\pi^0\pi^0\to\pi^0\pi^0\pi^0\gamma,\nonumber\\
&&e^+e^-\to\omega\eta\pi^0\to\eta\pi^0\pi^0\gamma,\nonumber\\
&&e^+e^- \to K_{S}K_{L}, \quad e^+e^- \to K_{S}K_{L}\pi^0,\nonumber\\
&&e^+e^- \to K_{S}K_{L}\pi^0\pi^0.\label{bkglist3}
\end{eqnarray}
Additional photons in events with three and four photons 
in the final state arise from splitting of the electromagnetic showers,
initial state radiation, and beam background.

To suppress background from the processes listed above,
a kinematic fits are performed to the hypotheses 
$ e^+e^-\to 3\gamma$, $ e^+e^-\to 5\gamma$, $e^+e^-\to\pi^0\pi^0\gamma$, and 
$e^+e^-\to\eta\pi^0\gamma$ with the requirement of energy and momentum 
balance in an event. For the two latter hypotheses, 
the additional constraints are imposed that the invariant masses 
of photon pairs are equal to the masses of the $\pi^0$ and $\eta$ mesons.
As a result of the kinematic fit, the energies and 
angles of photons are refined, and the $\chi^2$ of the 
proposed kinematic hypothesis is calculated. In the kinematics fits,
all possible combinations of photons are tested, and the combination with the
smallest $\chi^2$ value is retained. The following conditions are 
applied on the obtained $\chi^2$ values 
\begin{eqnarray}
&& \chi^2_{5\gamma}<30, \quad \chi^2_{\eta\pi^0\gamma}-\chi^2_{5\gamma}<10, \nonumber \\
&& \chi^2_{3\gamma}>50, \quad \chi^2_{\pi^0\pi^0\gamma}-\chi^2_{5\gamma}>80.
\label{signal}
\end{eqnarray}
498 events are selected using these conditions.
To estimate the background, along with the signal region determined by the 
conditions~(\ref{signal}), a control region is analyzed, for which the 
modified condition on the $\chi^2$ difference 
$10<\chi^2_{\eta\pi^0\gamma}-\chi^2_{5\gamma}<60 $ is used.

\section{\boldmath Fitting the $\pi^0\gamma$ mass and 
$\chi^2_{\eta\pi\gamma}-\chi^2_{5\gamma}$ distributions}
Selected events can be divided into four classes.
The first class ($\omega\eta$) contains events of the process 
\begin{equation}
\label{epgproc}
e^+e^-\to\omega\eta\to\eta\pi^0\gamma.
\end{equation}
The second class ($\rm res$-$\eta\pi\gamma$) contains events 
of the remaining hadron processes with the $\eta\pi^0\gamma$ final state:
\begin{equation}
\label{resepgproc1}
e^+e^-\to\rho\eta,\quad e^+e^-\to\phi\eta,\quad e^+e^-\to\phi\pi^0,
\end{equation}
\begin{equation}
\label{resepgproc2}
e^+e^-\to\omega\pi^0,\quad e^+e^-\to\rho\pi^0.
\end{equation}
The third class ($\rm rad$-$\eta\pi\gamma$) includes 
events from radiation decays of excited vector mesons, i.e. the processes
$e^+e^-\to a_{0}(980)\gamma$, $e^+e^-\to a_{0}(1450)\gamma$, 
and $e^+e^-\to a_{2}(1320)\gamma$. The last fourth class ($\rm bkg$)
are background events from the processes~(\ref{bkglist1}-\ref{bkglist3}).
The first three classes describe different intermediate mechanisms of the 
process $e^+e^-\to\eta\pi^0\gamma$.
\begin{figure*}
\resizebox{0.48\textwidth}{!}{\includegraphics{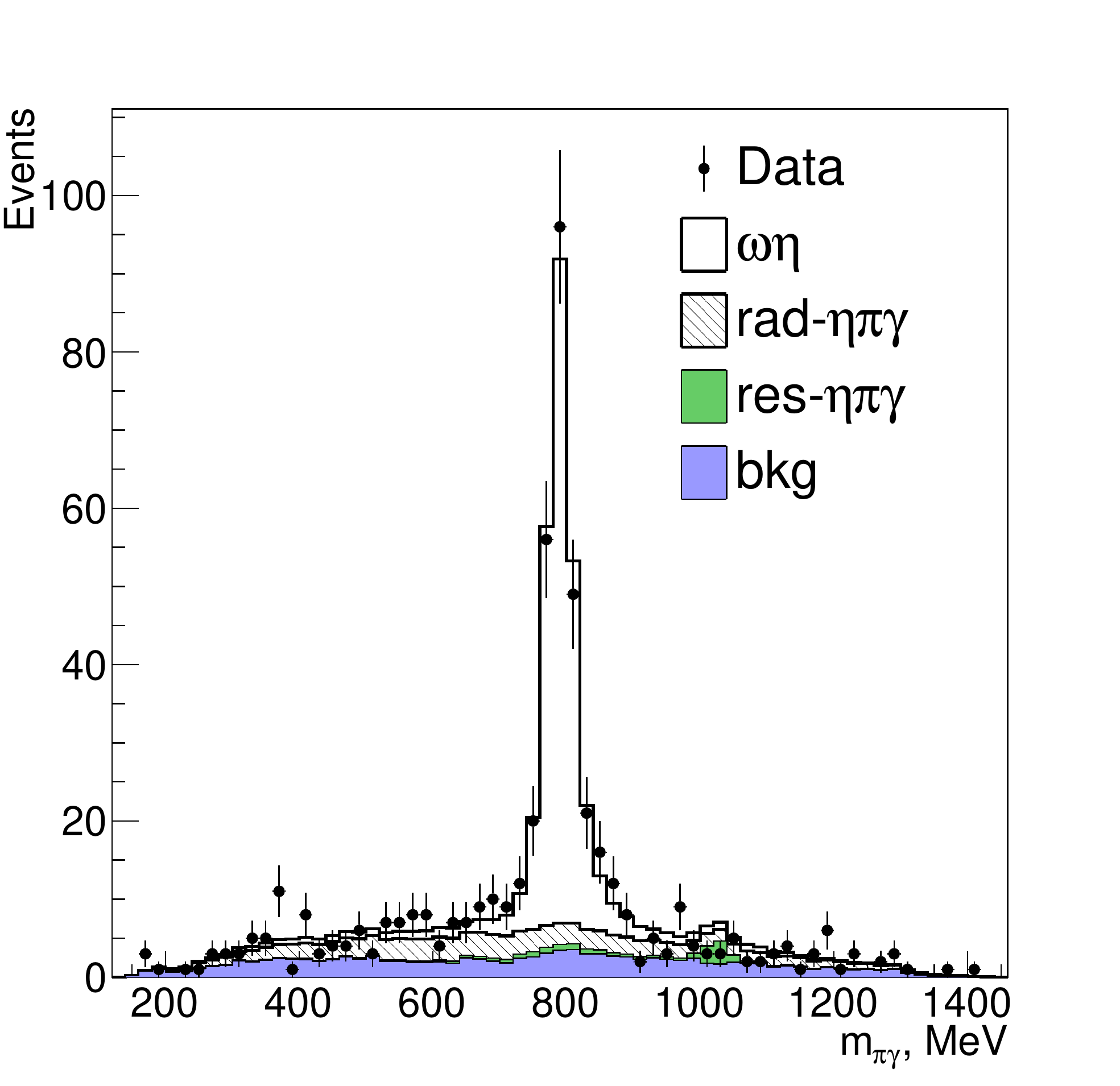}}\hfill
\resizebox{0.48\textwidth}{!}{\includegraphics{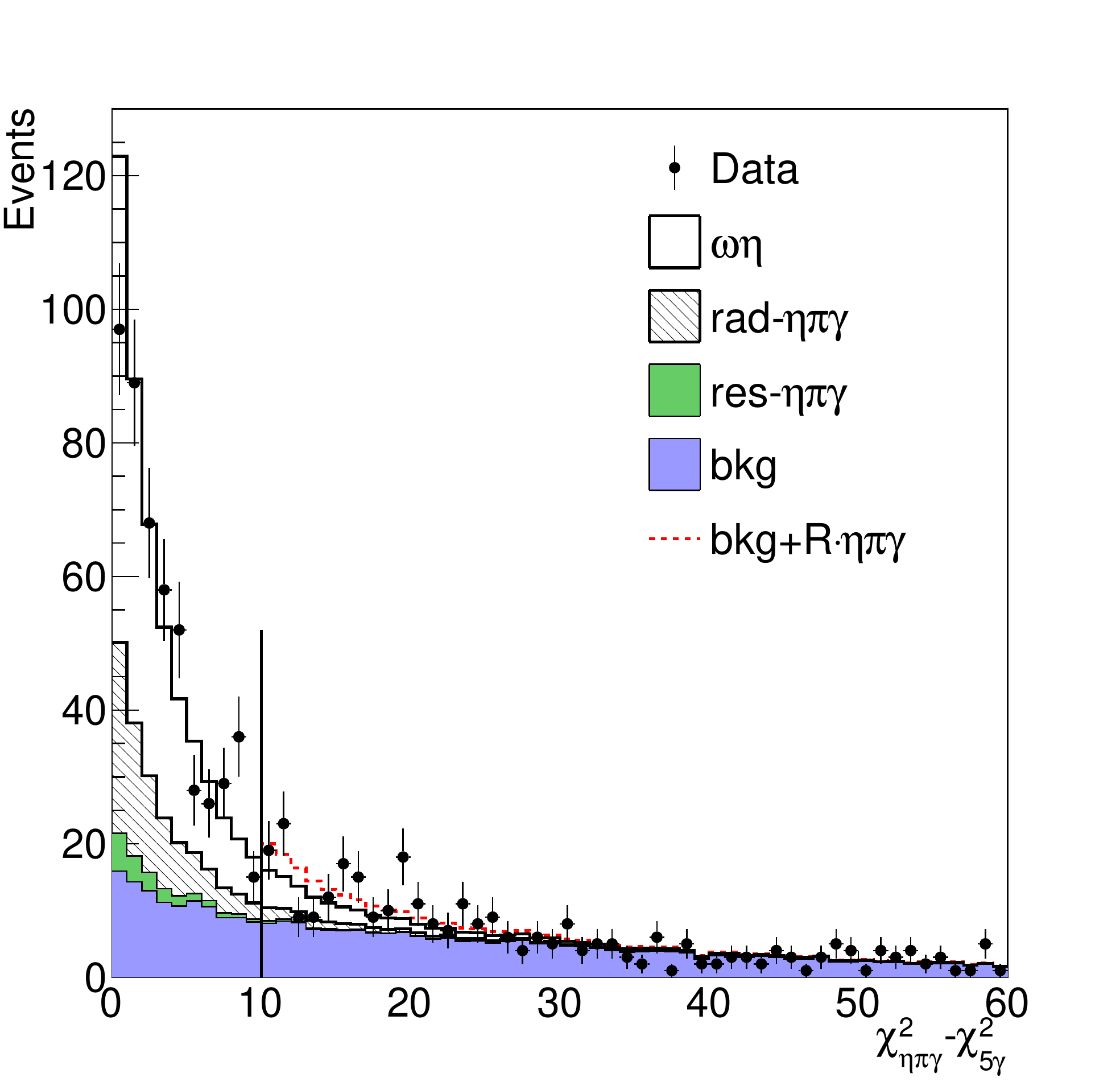}}
\caption{Left panel: The $\pi^0\gamma$ invariant mass distribution
for selected experimental events from the signal region 
$\chi^2_{\eta\pi\gamma}-\chi^2_{5\gamma}<10$ (points with error bars).
The histograms represents the results of the fit described in the
text. The distributions for the four event classes are shown cumulatively.
Right panel: The $\chi^2_{\eta\pi\gamma}-\chi^2_{5\gamma}$ distribution
for selected experimental events (points with error bars)
from the signal and control regions. 
The histograms shows cumulatively the simulated distributions for the
four event classes. The distributions are normalized to the number of 
events in the signal region. The dashed histogram
shows the result of the fit after the correction 
of the signal distribution described in the text. 
The vertical line separates the signal and control regions.
\label{FitResAll}}
\end{figure*}

The numbers of events in each class are determined using a combined fit to the
$\pi^0\gamma$ invariant mass distribution ($m_{\pi\gamma}$) for events from
the signal region ($\chi^2_{\eta\pi^0\gamma}-\chi^2_{5\gamma}<10$) and the 
$\chi^2_{\eta\pi^0\gamma}-\chi^2_{5\gamma}$ distribution for events from the
control region ($10<\chi^2_{\eta\pi^0\gamma}-\chi^2_{5\gamma}<60$).
The fit is performed using the maximum likelihood method.
The distributions of these parameters for all selected data events 
($\sqrt {s} = 1.05 $--2.00 GeV) are shown in Fig.~\ref{FitResAll}.


The distributions for the four classes of events used in the fit
are obtained by simulation. For the $\omega\eta$ class, we use
true particle parameters from the MC event generator to divide 
simulated $e^+e^-\to\omega\eta$ events into two subsets.
The first subset contains events, in which the kinematic fit chooses the 
correct photon combination for the candidate $\eta$ meson. The second subset
contains events with wrong combinations, which are responsible 
for the long tails in the $m_{\pi\gamma}$ distribution. This
distribution is described by a sum of distributions for the two subset:
$(1-k_{\rm wide})H_1+k_{\rm wide}H_2$. The fraction of events of the
second subset $k_{\rm wide}$ determined using simulation decrease from 0.17
near the $\omega\eta$  threshold to 0.12 at $\sqrt{s}=2$ GeV.

To take into account the imperfect simulation of the $\omega$ meson line shape,
a mass shift $\Delta m$ and Gaussian smearing with the dispersion
$\Delta\sigma^2$ are introduced into the $\pi^0\gamma$ mass spectrum. 
These parameters are determined from comparison of the $\omega$ meson
peak position and width in data and simulation from the range
$\sqrt {s} = 1.41$--1.80~GeV. They are found to be $\Delta m = 4.0 \pm 1.6$~MeV
and $\Delta\sigma^2 = -48\pm48$~MeV$^2$.
We do not expect that the
$m_{\pi\gamma}$ resolution in data is better than that in simulation.
Therefore, we regard the negative $\Delta\sigma^2$ value as a statistical
fluctuation and do not correct the $\pi^0\gamma$ mass resolution.

To estimate how accurate the simulation reproduces the value of the parameter
$k_{\rm wide}$, $e^+e^-\to\omega\pi^0$ events are studied in the energy range
$\sqrt{s}=1.05$--1.6 GeV, where the background to this process is negligible.
For this process, the fraction of events in the long distance tails of 
the $m_{\pi\gamma}$ distribution is reproduced by simulation
with a statistical accuracy of 10\%. This number is taken as a measure
of the systematic uncertainty of the $k_{\rm wide}$ coefficient.

To describe the $\rm res$-$\eta\pi\gamma$ contribution, 
the simulation of the processes~(\ref{resepgproc1}) is used.
The distributions obtained from simulation are normalized to expected number
of events and summed up. To calculate the expected number of events, we used
the cross sections for the processes~(\ref{resepgproc1}) measured in 
Refs.~\cite{BaBar_rhoeta,CMD_phieta,BaBar_phieta,BaBar_phipi}
and the $\rho\to\pi^0\gamma$, $\phi\to\pi^0\gamma$, and $\eta \gamma$
branching ratios~\cite{pdg}. The contribution of the 
processes~(\ref{resepgproc2}) estimated using Ref.~\cite{SND_ompi,SND_rhopi} 
is found to be negligibly small. The total 
$\rm res$-$\eta\pi\gamma$ contribution is calculated to be 
$N_{\rm res} = 19.6 \pm 1.3$.

To describe the $\rm rad$-$\eta\pi\gamma$ contribution,
the simulation of the processes $e^+e^-\to a_{0}(1450)\gamma$ 
and $e^+e^-\to a_{2}(1320)\gamma$ is used. The distributions for these 
processes are summed with the weights $1-\alpha$ and $\alpha$, 
respectively, where the coefficient $\alpha = 0.22 \pm 0.21$. 
The choice of the value of this coefficient is discussed below 
in Sec.~\ref{raddecmodel}. It should be noted that the 
fit results very weakly depend on the value of $\alpha$.

The $m_{\pi\gamma}$ and 
$\chi^2_{\eta\pi^0\gamma}-\chi^2_{5\gamma}$ distributions 
for background  are calculated using simulation of the 
processes~(\ref{bkglist1}-\ref{bkglist3}).
The distributions obtained for each process are normalized
to the expected number of events and summed up. Only the shape of
the distribution is used in the fit. The total number of background events
is a free fit parameter.

\begin{figure*}
\center
\resizebox{0.48\textwidth}{!}{\includegraphics{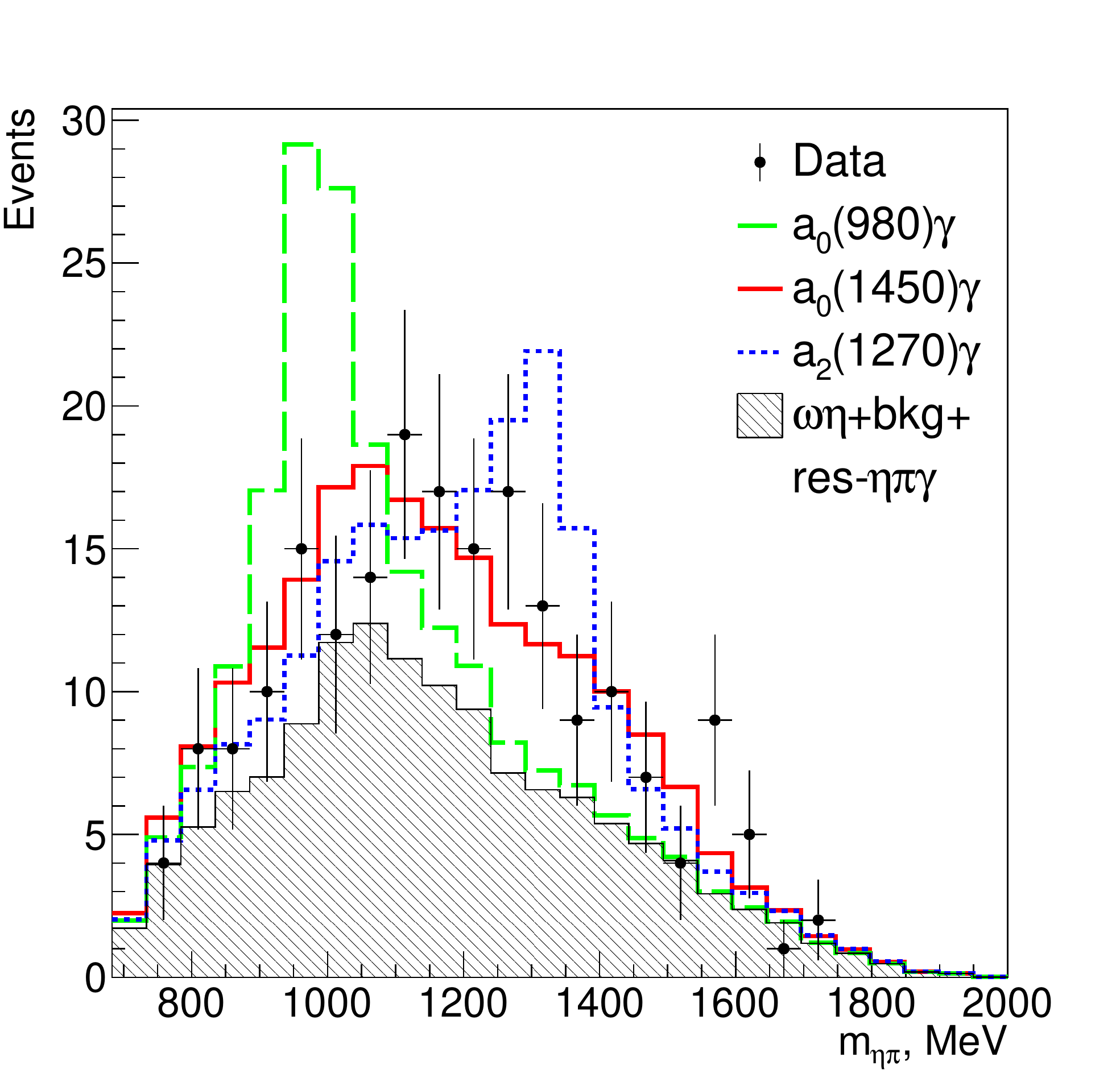}}\hfill
\resizebox{0.48\textwidth}{!}{\includegraphics{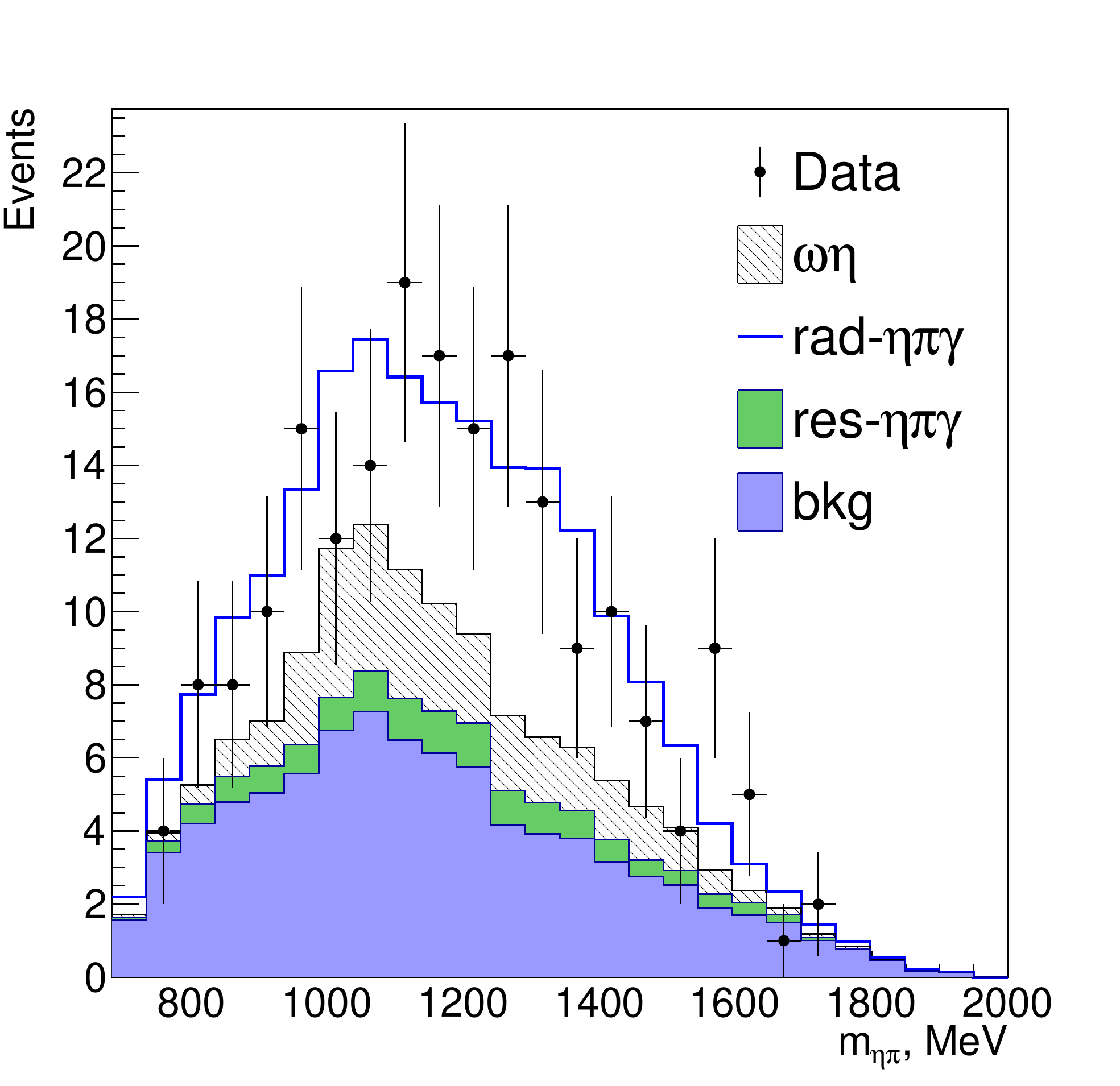}}
\caption{The $\eta\pi^0$ invariant mass distribution
for data (points with error bars) from the interval $\sqrt{s}=1.05$--2.00 GeV.
Events from the $\omega$ meson peak ($700<m_{\pi\gamma}<900$ MeV) are rejected.
Left panel: The shaded histogram shows the total distribution 
for the background and the processes (\ref{epgproc}) and (\ref{resepgproc1}).
The solid, dotted and dashed histograms show the total distribution 
for events of all four classes, in which the distribution for the class 
$\rm rad$-$\eta\pi\gamma$ are calculated in the models 
$a_{0}(1450)\gamma$, $a_{2}(1320)\gamma$, and $a_{0}(980)\gamma$, respectively.
Right panel: The contributions of the four classes of events are shown
cumulatively. The distribution for the class $\rm rad$-$\eta\pi\gamma$
is obtained from the fit to the two-dimensional distribution of 
$m_{\pi\gamma}$ versus $m_{\eta\pi}$ as described in the text.
\label{FitResMepAll}}
\end{figure*}
The following relations between the numbers of signal and background events 
in the signal and control regions are used in the fit:
$N_{\rm bkg}=k_{\rm bkg}N^{\rm C}_{\rm bkg}$,
$N_{\eta\pi\gamma}^{\rm C}=k_{\rm sig}N_{\eta\pi\gamma}$,
where $N_{\eta\pi\gamma}$ and $N_{\rm bkg}$ are the numbers of
$e^+e^- \to \eta\pi^0\gamma$ and background events in the signal region,
respectively, and $N_{\eta\pi\gamma}^{\rm C}$ and $N^{\rm C}_{\rm bkg}$ are
the same numbers in the control region.
The coefficients $k_{\rm sig}$ and $k_{\rm bkg} $ are calculated using
simulation in each of the 13 energy intervals. The uncertainty of 
$k_{\rm bkg}$ is estimated by varying the cross sections of the background 
processes~(\ref{bkglist1}-\ref{bkglist3}) within their errors. The value of
this coefficient averaged over the energy range $\sqrt{s} = 1.05$--2.00~GeV  
is $k_{\rm bkg} = 0.53 \pm 0.01$. The coefficient $k_{\rm sig}$ obtained
from simulation is corrected to take into account the difference in the 
signal $\chi^2_{\eta\pi^0\gamma}-\chi^2_{5\gamma}$ distributions
in the data and simulation. To do this, data from the energy range
$\sqrt{s} = 1.41$--1.80~GeV are used, where the cross section of 
the process $e^+e^-\to\omega\eta$ is maximal. The mass spectrum 
of $\pi^0\gamma$ is analyzed and the number of data events in the 
$\omega$ meson peak is determined for the signal 
($N^{\rm data}_{\rm S}$) and control ($N^{\rm data}_{\rm C}$) regions.
The numbers of simulated  $e^+e^-\to\omega\eta$ events in the signal
($N^{\rm MC}_{\rm S}$) and control ($N^{\rm MC}_{\rm C}$) regions
are also determined. The double ratio 
$R = (N^{\rm data}_{\rm C}N^{\rm MC}_{\rm S})/(N^{\rm data}_{\rm S}
N^{\rm MC}_{\rm C})=1.5 \pm 0.3$ is used to correct the 
coefficient $k_{\rm sig}$. With this correction, this coefficient
averaged over the full energy range is $k_{\rm sig}=0.22 \pm 0.04$.
The simulation shows that the difference between the 
$\chi^2_{\eta\pi^0\gamma}-\chi^2_{5\gamma}$ distributions for different
intermediate mechanisms of the process $e^+e^-\to\eta\pi^0\gamma$ is 
very small.

To evaluate the systematic uncertainty associated with the procedure of 
determining the number of events in each class, the nuisance parameters 
corresponding to $N_{\rm res}$, $\Delta m$, $\Delta\sigma^2$, and the 
coefficients $k_{\rm wide}$, $ k_{\rm sig} $, and $k_{\rm bkg}$ are introduced 
into the likelihood function with Gaussian constraints. The fit results are
represented by the solid histogram in Fig.~\ref{FitResAll} (left) for 
the $m_{\pi\gamma}$ distributions, and by the dashed histogram for the
$\chi^2_{\eta\pi^0\gamma}-\chi^2_{5\gamma}$ distribution in 
Fig.~\ref{FitResAll} (right). The latter take into account the scale factor
$R$ for the signal events described above, while the solid histogram 
in Fig.~\ref{FitResAll} shows the uncorrected distribution normalized to the
number of events in the signal region. 
The following numbers of events of three classes are obtained in
the fit:
\begin{eqnarray}
&&N_{\omega\eta} = 267 \pm 20, \quad N_{\rm rad} = 101 \pm 21,\nonumber\\
&& N_{\rm bkg} = 113 \pm 10.\label{NevtFit}
\end{eqnarray}
The correlations between these numbers are not large.
The correlation coefficients are $-0.25$ between  $N_{\omega\eta}$ and $N_{rad}$, 
$-0.51$ between  $N_{bkg}$ and $N_{rad}$, and $-0.16$ 
between  $N_{\omega\eta}$ and $N_{bkg}$.
The total contribution from the background 
processes~(\ref{bkglist1}-\ref{bkglist3}) calculated using simulation is
$N_{\rm bkg}^{\rm calc} = 123$ and agrees well with the fit result.

The obtained numbers of $\omega\eta$ and $\rm rad$-$\eta\pi\gamma$ events in
13 energy intervals are listed in 
Tables~\ref{omegatab} and \ref{nonomegatab}, respectively. 
Table~\ref{nonomegatab} also shows the distribution over the energy intervals
for background events and events of the class $\rm res$-$\eta\pi\gamma$.

\section{\boldmath
Model selection for the $\rm rad$-$\eta\pi\gamma$ event class
\label{raddecmodel}}
As it is mentioned in the previous section, the following three processes
may contribute to the $\rm rad$-$\eta\pi\gamma$ class in the energy 
range 1.05--2.00 GeV:
\begin{eqnarray}
&&e^+e^-\to a_{0}(980)\gamma, \nonumber\\
&&e^+e^-\to a_{0}(1450)\gamma,\nonumber\\
&&e^+e^-\to a_{2}(1320)\gamma \label{radmods}
\end{eqnarray}
Figure~\ref{FitResMepAll}(left) shows the $\eta\pi^0$ invariant mass
($m_{\eta\pi}$) spectrum for selected events from the interval
$\sqrt{s}=1.05$--2.00 GeV. Events from the $\omega$ meson peak 
($700 <m_{\pi\gamma}<900$ MeV) are rejected.
The shaded histogram shows the total distribution for the background
and the processes (\ref{epgproc}) and (\ref{resepgproc1}).
The solid, dotted and dashed histograms show the total distribution 
for events of all four classes, in which the distribution for the class 
$\rm rad$-$\eta\pi\gamma$ is calculated in the models 
$a_{0}(1450)\gamma$, $a_{2}(1320)\gamma$, and $a_{0}(980)\gamma$, respectively.
The distributions are normalized to the numbers of events (\ref {NevtFit}) 
found in the previous section. It is seen the observed $\eta\pi^0$ mass
spectrum is best described by the model $e^+e^-\to a_{0}(1450)\gamma$.

The following model is tested to describe distributions 
for $\rm rad$-$\eta\pi\gamma$ events:
\begin{eqnarray}
P_{rad} &=& \alpha P_{a_2(1320)\gamma} + \beta P_{a_0(980)\gamma}\nonumber\\ 
&+& (1-\alpha-\beta) P_{a_0(1450)\gamma},\label{Prad}
\end{eqnarray}
where $P$ is, for example, a two-dimensional distribution of  
$m_{\pi\gamma}$ versus $m_{\eta\pi}$. This simple model 
does not take into account the interference between the three 
intermediate mechanisms, but can be used to estimate the model uncertainty
of the efficiency for  $\rm rad$-$\eta\pi\gamma$ events.

The data from the range $\sqrt{s}=1.05$--2.00 GeV are fitted as
described in the previous section. To increase the sensitivity 
to the model of the intermediate states~(\ref{Prad}),
the $m_{\pi\gamma}$ distribution for the signal region 
is replaced by the two-dimensional distribution of
$m_{\pi\gamma}$ versus $m_{\eta\pi}$. The parameters $\alpha$ 
and $\beta$ are determined from the fit with the constraints that 
$\alpha$, $\beta$ and $\alpha+\beta$ vary from 0 to 1.
The following values of these parameters are obtained:
$\alpha = 0.22\pm 0.21$ and $\beta = 0.00^{+0.08}$. 
The $m_{\eta\pi}$ distribution obtained using simulation with these 
parameters is shown in Fig.~\ref{FitResMepAll}(right).

To evaluate the significance of the $\rm rad$-$\eta\pi\gamma$ signal,
we compare the values of likelihood function for the fit described 
above ($L_1$) and the fit with the $N_{rad}\equiv 0$ ($L_0$).
Taking into account that the numbers of parameters in these 
two fit differ by three, from the difference 
$-2(\ln{L_0}-\ln{L_1})=39.5$, we obtain that the significance of the observed
$\rm rad$-$\eta\pi\gamma$ signal (including the systematic uncertainty)
is 5.6 $\sigma$. 

\section{Detection efficiency and radiative corrections}
\begin{figure}
\centering
\resizebox{0.45\textwidth}{!}{\includegraphics{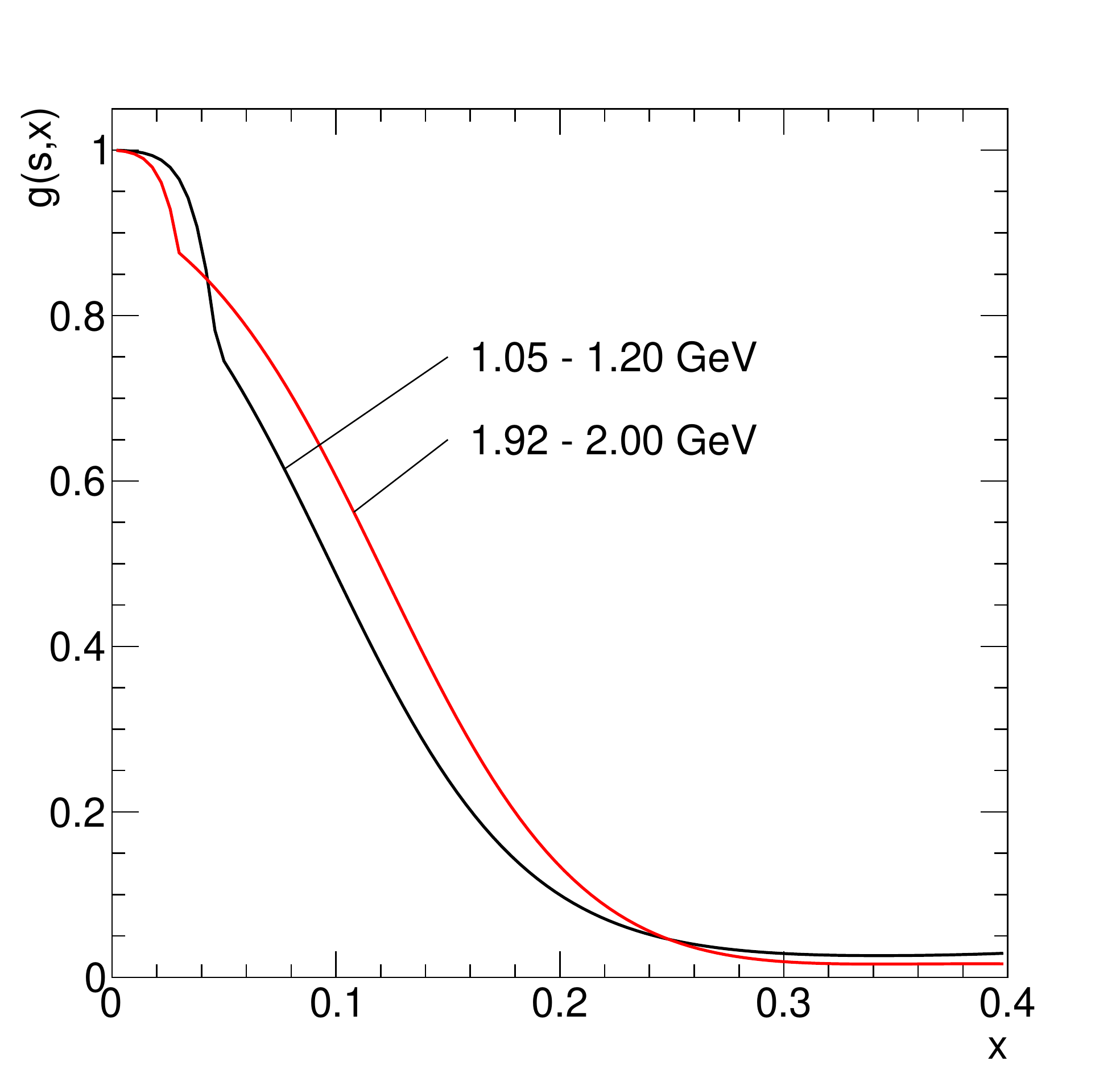}}
\caption{The detection efficiency for 
$e^+e^- \to \omega \eta \to \eta\pi^0\gamma$ events as a function of the 
normalized ISR photon energy for two energy intervals.
\label{EffRad}}
\end{figure}
The detection efficiency is determined using simulation
as a ratio of the number of selected events to the number of generated events.
It includes both the detector acceptance and selection efficiency.
The simulation of the process $e^+e^- \to \eta\pi^0\gamma$
takes into account initial state radiation (ISR).
The detection efficiency for the process under study 
is calculated as a function of two parameters, $\sqrt{s}$ and normalized
energy of the ISR photon, 
$x=2E_\gamma/\sqrt{s}$, and is parametrized as follows 
$\varepsilon_r(s,x) = \varepsilon(s) g(s,x)$, where 
$\varepsilon(s) \equiv \varepsilon_r(s, 0)$. The function $g(s,x)$ 
weakly depends on $\sqrt{s}$ and on the intermediate mechanism 
of the process $e^+e^- \to \eta\pi^0\gamma$. Its dependence 
on $x$ is shown in Fig.~\ref{EffRad}. The shape of this dependence is 
determined by two effects. The sharp decrease of the efficiency 
at $x$ corresponding to $E_{\gamma}=20$--30~MeV is due to the condition 
that an event contain exactly five photon. This condition rejects events
with the ISR photon emitted at a large angle. Photons are reconstructed 
if their energy deposition in the calorimeter exceeds 20 MeV.  A further 
decrease of the efficiency is due to the requirement of 
energy-momentum balance in an event ($\chi^2_{5\gamma}<30$).

The visible cross section of the process $e^+e^- \to \eta\pi^0\gamma$,
which is defined as $\sigma_{\rm vis}=N/L$, where $N$ is the number of
selected events of the process under study and $L$ is the integrated 
luminosity, is related to the Born cross section $\sigma(s)$ as follows:
\begin{equation}
\sigma_{\rm vis}(s) = \int
\limits_{0}^{x_{max}}\varepsilon_r(s,x)F(x,s)
\sigma(s (1-x))dx,
 \label{viscrs}
\end{equation}
where $F(x,E)$ is a so-called radiator function~\cite{FadinRad} describing 
the probability to emit extra photons with the total energy $x\sqrt{s}/2$
from the initial state, and $x_{max}=1-(m_\eta+m_{\pi^0})^2/s$.
The formula (\ref{viscrs}) can be rewritten in the conventional form:
\begin{equation}
\label{viscrs2} \sigma_{\rm vis}(s) =
\varepsilon(s)\sigma(s)(1+\delta(s)),
\end{equation}
where $\delta(s)$ is the radiative correction.

Inaccuracy in simulation of the detector response for photons 
leads to systematic uncertainty in the detection efficiency determined 
using the simulation. To evaluate the efficiency corrections associated 
with the selection criteria, we change the boundaries of the conditions:
on $\chi^2_{5\gamma}$ from 30 to 60,  
in $\chi^2_{\eta\pi^0\gamma} - \chi^2_{5\gamma}$ from 10 to 60, and on 
$\chi^2_{\pi^0\pi^0\gamma}-\chi^2_{5\gamma}$ from 80 to 10, and remove
the condition on $\chi^2_{3\gamma}$. The relative change 
in the $e^+e^-\to\omega\eta$ cross section after
loosening the selection condition is taken as a
correction for the detection efficiency, while the uncertainty of this 
correction is added to its systematic uncertainty.
The total efficiency correction due to these conditions is $-(5.7 \pm 6.1)\%$.
The minus sign means that the efficiency in the data is less than that 
in the simulation. The efficiency correction for the 
condition $N_\gamma=5$ is determined using five-photon 
events of the process $e^+e^-\to\omega\pi^0$, the cross section for 
which can be measured with the condition $N_\gamma\geq 5$, and 
found to be $-(0.4\pm 0.2)\%$.

In SND, photons converting into a $e^+e^-$ pair in the material
before the drift chamber, produce a charged track. Such events do not pass 
the selection criteria.
Since the process under study and the process used for normalization contain 
different numbers of photons in the final state, improper simulation of 
the photon conversion lead to a shift in the measured cross section. 
The photon conversion probability is measured using events of the 
process $e^+e^-\to\gamma\gamma$. The corresponding efficiency correction
is found to be $(-0.79 \pm 0.02)\%$. Thus, the total 
correction for the detection efficiency is $(-6.9\pm 6.1)\%$.

The detection efficiency for the class $\rm rad$-$\eta\pi\gamma$ is
calculated in the model~(\ref{Prad}) with the coefficients 
$\alpha = 0.22\pm 0.21$ and $\beta=0.00^{+0.08}$. The model uncertainty of 
the efficiency is determined by varying the coefficients $\alpha$ and 
$\beta$ within their errors. It does not exceed 3\%.

\section{Fitting the measured cross sections}
To calculate the radiative correction and determine the Born cross section,
the energy dependence of the visible cross section is fitted by 
Eq.~(\ref{viscrs}). The uncertainty on the radiative correction is 
estimated by varying the fitted parameters within their errors. 
The energy dependence of the Born cross section is 
parametrized by a sum of contributions of two vector resonances:
of the as follows
\begin{eqnarray}
\sigma(s) &=&
\frac{12\pi}{s^{3/2}} \left| \sqrt{\frac{B_{V^{\prime}}}{P_{f}(m_{V^{\prime}})}}
\frac{m_{V^{\prime}}^{3/2}\Gamma_{V^{\prime}}}{D_{V^{\prime}}}\right.\nonumber\\
& + & \left.
\sqrt{\frac{B_{V^{\prime\prime}}}{P_{f}(m_{V^{\prime\prime}})}}
\frac{m_{V^{\prime\prime}}^{3/2}\Gamma_{V^{\prime\prime}}}{D_{V^{\prime\prime}}}
e^{i\varphi}
\right|^{2} P_{f}(s),\label{crsborn}
\end{eqnarray}
where $m_V$ and $\Gamma_V$ are mass and width of the resonance, 
$V$ ($V=V^{\prime}$ or $V^{\prime\prime}$), 
$D_{V}= s - m_{V}^2 +i \sqrt{s}~\Gamma_{V}$,
$B_{V} = B(V \to e^+e^-)B(V \to f)$ is the product of the branching 
fractions of $V$ to $e^+e^-$ and to the final state $f$, 
$P_{f}(s)$ is the phase-space factor.
\begin{figure}
\centering
\resizebox{0.48\textwidth}{!}{\includegraphics{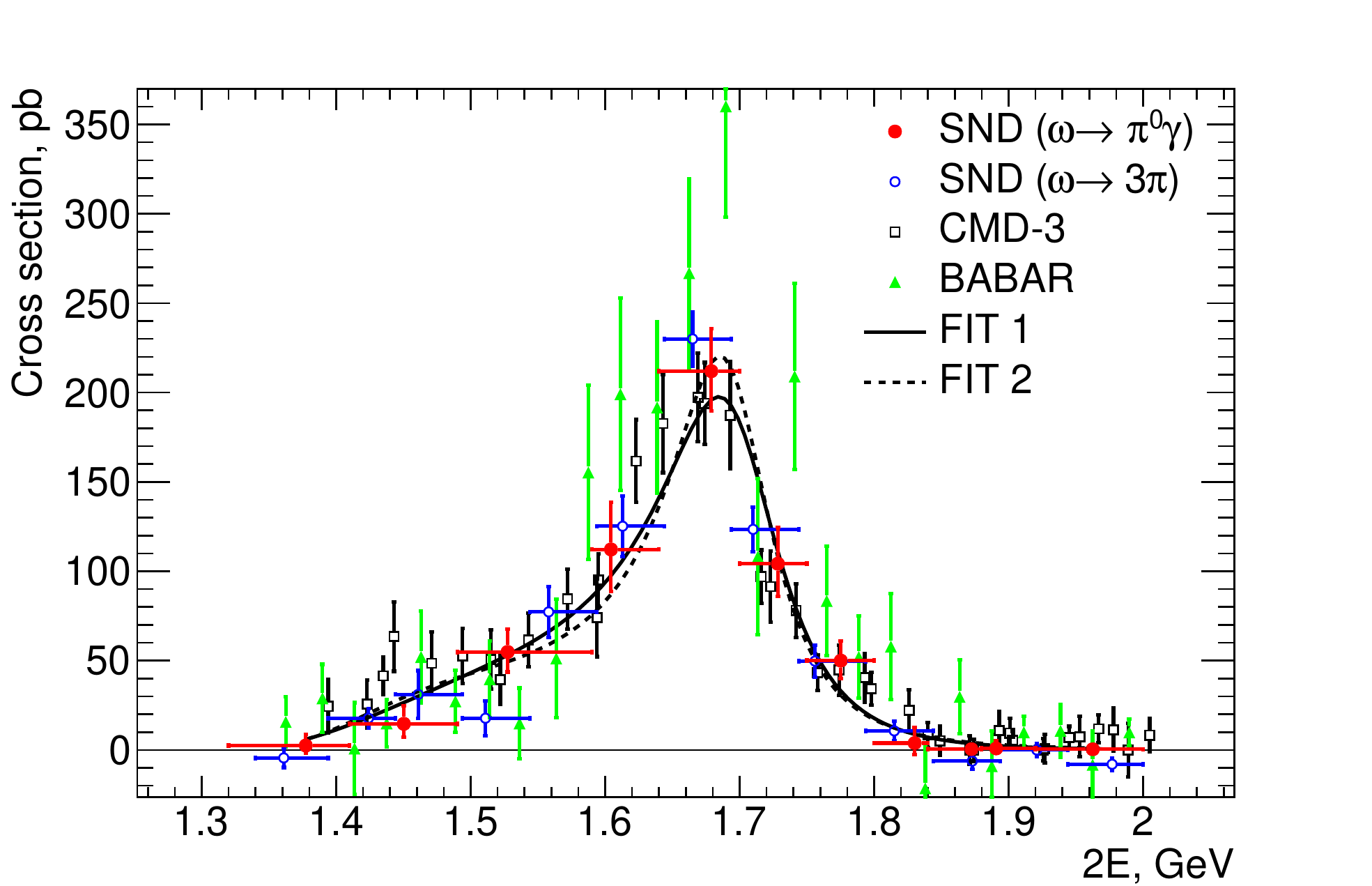}}\hfill
\caption{The energy dependence of the 
 $e^+e^-\to \omega\eta \to \eta\pi^{0}\gamma$ Born cross section measured 
in this work (filled circles). For comparison, the SND~\cite{eta3pi-snd}
(open circles), CMD-3~\cite{eta3pi-cmd} (squares), and BABAR~\cite{ometa-babar}
(triangles) measurements of the $e^+e^-\to \omega\eta$ cross section  
in the decay mode $\omega\to \pi^+\pi^-\pi^0$ are shown. 
These data are multiplied by the branching fraction 
$B(\omega\to\pi^0\gamma)$~\cite{SND_pig}. The curves show
the results of the fit described in the text.
\label{OmeCrsPic}}
\end{figure}

In the fit to the $e^+e^-\to\omega\eta\to\eta\pi^{0}\gamma$ data,
the first term in Eq.~(\ref{crsborn}) is assumed to be 
the contribution of the $\omega(1420)$ resonance, while the second
effectively describes the contributions of two resonances, $\omega(1650)$
and $\phi(1680)$, which have close masses.  The phase-space factor
is calculated in the approximation of narrow $\omega$ meson width:
$P_{f}(s) = q_{\omega}^3(s)$, where $q_{\omega}(s)$ is the $\omega$ meson 
momentum in the reaction $e^+e^-\to\omega\eta$. The phase $\varphi$
between the first and second terms is chosen equal to $\pi$.
The fit is performed in two variants. In the first, the $V^{\prime}$ 
mass is fixed at the Particle Data Group (PDG) 
value~\cite{pdg} $m_V^{\prime} = 1420$ MeV, while $\Gamma_{V^{\prime}}$,
$m_{V^{\prime\prime}}, \Gamma_{V^{\prime\prime}}, B_{V^{\prime}}$
and $B_{V^{\prime\prime}}$ are free parameters. In the second,
we follow the works~\cite{eta3pi-snd,eta3pi-cmd} and fix
also the $\omega(1420)$ width at $\Gamma_{V^{\prime}}=220$~MeV~\cite{pdg}.
The fitted curves are shown in Fig.~\ref{OmeCrsPic}.
The obtained values of the fit parameters are listed in
Table~\ref{CrsFitRes}. 
They are in good agreement with the results of the previous 
works~\cite{eta3pi-snd,eta3pi-cmd}.

The first variant of the fit has a better $\chi^2$. It is used to calculate
the radiative corrections. The obtained 
values of the radiative correction and the Born cross section are listed in
Table~\ref{omegatab}. In the first column of the table we list
the boundaries of the energy interval and the weighted average energy, 
which is calculated as
$\sum_i\sqrt{s_i}\sigma_{{\rm vis},i}L_i/\sum_i\sigma_{{\rm vis},i}L_i$,
where the sum is taken over the energy points entering in the energy
interval, and the visible cross section is calculated using Eqs.~(\ref{viscrs})
and (\ref{crsborn}) with the parameters obtained in the fit. For the cross
section, the statistical and systematic errors are quoted.
\begin{table*}[p]
\centering
\caption{The parameters obtained in the fits to the 
$e^+e^-\to\omega\eta$ and  $e^+e^-\to {\rm rad}\mbox{-}\eta\pi\gamma$
cross sections.  $B_{V} = B(V\to e^+e^-)B(V\to \omega\eta)$ for $\omega\eta$,
and $B_{V} = B(V\to e^+e^-)B(V\to \eta\pi\gamma)$ for 
$\rm rad$-$\eta\pi\gamma$. For the $\omega\eta$ final state the results 
for the two variants of the fit are listed in comparison with the results of
the SND~\cite{eta3pi-snd} and CMD-3~\cite{eta3pi-cmd} experiments obtained
in the $\omega\to\pi^+\pi^-\pi^0$ decay mode.
\label{CrsFitRes}}
\begin{tabular}{|c|c|c|c|c|c|}
\hline
& $\omega\eta$(1) & $\omega\eta$(2) & SND($3\pi$) & CMD($3\pi$)&$\rm rad$-$\eta\pi\gamma$\\  \hline 
$m_{V^{\prime}}$, MeV & $\equiv 1420$ & $\equiv 1420$ &$\equiv 1420$ & $\equiv 1420$ & $1415 \pm 52$ \\
$\Gamma_{V^{\prime}}$, MeV & $440\pm125$ & $\equiv 220$ & $\equiv 220$ & $\equiv 220$& $247\pm 81$ \\
$B_{V^{\prime}} \times 10^{8}$ & $2.5 \pm 0.6$ & $3.0\pm0.6$ & $2.1^{+1.0}_{-0.8}$ & $3.2\pm0.6$& $0.04 \pm 0.02$ \\
$m_{V^{\prime\prime}}$, MeV & $1698 \pm 10 $ & $1694\pm 9$ & $1673^{+6}_{-7}$ & $1679\pm5$ & --- \\
$\Gamma_{V^{\prime\prime}}$, MeV & $110 \pm 16 $ & $94 \pm 13$ & $95\pm11$ & $121\pm9$& --- \\
$B_{V^{\prime\prime}} \times 10^{7}$ & $6.4 \pm 0.9 $ & $5.4 \pm 0.6$ & $5.62^{+0.45}_{-0.42}$ & $4.7\pm0.3$& --- \\
$\chi^2/n.d.f.$ & 8.6/6 & 12.4/7 & 10.5/9 & 23/35 &10.8/10\\
\hline
\end{tabular}
\end{table*}

\begin{table*}[p]
\centering
\caption{\normalsize 
The weighted average energy ($\sqrt{s}$) for the interval indicated in 
parentheses, integrated luminosity ($L$), detection efficiency 
($\varepsilon_{\omega\eta}$),
number of selected events ($N_{\omega\eta}$), radiative correction
($1+\delta$),
and Born cross section ($\sigma_{\omega\eta}$) for the process 
$e^+e^-\to\omega\eta\to\eta\pi^0\gamma$. For the cross section,
the first error is statistical, and the second is systematic.
\label{omegatab}}
\begin{tabular}{|c|c|c|c|c|c|}
\hline
$\sqrt{s}$, GeV & L, pb$^{-1}$ & $\varepsilon_{\omega\eta}$, \% & $N_{\omega\eta}$ & $1+\delta$ & 
$\sigma_{\omega\eta}$, pb  \\  \hline 
1.38(1.32-1.41) &  6.26 & 9.36 & $1.2^{+2.9}_{-2.0}$ & 0.801 & $2.5^{+6.2}_{-4.2}$ $\pm$ 1.4\\ 
1.45(1.41-1.49) &  4.00 & 9.35 & $4.6^{+3.1}_{-2.3}$ & 0.841 & $14.6^{+9.9}_{-7.5}$ $\pm$ 3.1\\ 
1.53(1.49-1.59) &  6.52 & 9.35 & 28.7 $\pm$ 6.2 & 0.857 & 55 $\pm$ 12 $\pm$ 8\\ 
1.60(1.59-1.64) &  3.68 & 8.89 & 31.3 $\pm$ 7.0 & 0.852 & 112 $\pm$ 25 $\pm$ 12\\ 
1.68(1.64-1.70) &  6.20 & 9.40 & 105.8 $\pm$ 11.5 & 0.856 & 212 $\pm$ 23 $\pm$ 18\\ 
1.73(1.70-1.75) &  3.60 & 10.28 & 37.0 $\pm$ 6.9 & 0.964 & 104 $\pm$ 19 $\pm$ 10\\ 
1.78(1.75-1.80) &  7.00 & 9.91 & 39.3 $\pm$ 7.3 & 1.153 & 49 $\pm$ 9 $\pm$ 7\\ 
1.83(1.80-1.84) &  3.99 & 10.07 & $2.1^{+3.5}_{-2.6}$ & 1.405 & $3.7^{+8.7}_{-6.5}$ $\pm$ 0.9\\ 
1.87(1.84-1.88) & 15.64 & 8.52 & $0.9^{+4.1}_{-3.1}$ & 1.523 & $0.4^{+3.1}_{-2.3}$ $\pm$ 0.2\\ 
1.89(1.88-1.92) & 11.43 & 8.92 & $1.3^{+4.3}_{-3.0}$ & 1.526 & $0.9^{+4.2}_{-2.9}$ $\pm$ 0.4\\ 
1.96(1.92-2.00) & 12.80 & 8.55 & $0.7^{+2.6}_{-1.8}$ & 1.389 & $0.5^{+2.4}_{-1.6}$ $\pm$ 0.2\\ 
\hline
\end{tabular}
\end{table*}
\begin{table*}[p]
\center
\caption{The main sources of the systematic uncertainty on the measured 
$e^+e^-\to\omega\eta$ and $e^+e^-\to {\rm rad}\mbox{-}\eta\pi^0\gamma$
cross sections, and the total $e^+e^-\to \eta\pi^0\gamma$ cross section.
\label{syser}}
\begin{tabular}{|c|c|c|c|c|}
\hline
Source& $\omega\eta$(1.64-1.70 GeV) & $\omega\eta$(1.84-2.00 GeV)& 
$\rm rad$-$\eta\pi\gamma$ & $\eta\pi^0\gamma$ \\  \hline 
Luminosity & 2\% & 2\% & 2\% & 2\% \\
Selection conditions & 6\% & 6\% & 6\% & 6\% \\
Determination of the number of signal events & 2\% & 20-27\% & 4-35\% & 1-15\% \\
Efficiency model dependence & --- & --- & 2-3\% & 0.5-3\% \\
Interference with the $\rho\eta$ final state & 5\% & 24-37\% & --- & --- \\
Radiative correction & 1\%& 13-27\% & 1-3\% & 1-6\%\\ \hline
Total & 8\% & 40-50\% & 8-35\% & 6-18\% \\
\hline
\end{tabular}
\end{table*}

\begin{table*}[p]
\center
\caption{\normalsize
The weighted average energy ($\sqrt{s}$) for the interval indicated in 
parentheses, integrated luminosity ($L$), detection efficiency
($\varepsilon_{\rm rad}$), number of selected events ($N_{\rm rad}$) for 
the process $e^+e^-\to {\rm rad}\mbox{-}\eta\pi^0\gamma$, number of events
from the processes $e^+e^-\to\rho\eta$, $\phi\eta$, $\phi\pi$ ($N_{\rm res}$),
number of background events ($N_{\rm bkg}$), radiative correction
($1+\delta$), and Born cross section for the process 
$e^+e^-\to {\rm rad}\mbox{-}\eta\pi^0\gamma$. For the cross section,
the first error is statistical, the second is systematic.
\label{nonomegatab}}
\begin{tabular}{|c|c|c|c|c|c|c|c|}
\hline
$\sqrt{s}$, GeV & $L$, pb$^{-1}$ & $\varepsilon_{rad}$, \% & $N_{rad}$ &
$N_{\rm res}$ & $N_{\rm bkg}$ & $1+\delta$ &
$\sigma_{\rm rad}$, pb \\  \hline 
1.15(1.05-1.20) &  4.86 & 5.68 & $2.0^{+2.6}_{-2.0}$ & 0.0 $\pm$ 0.1 & 1.7 $\pm$ 1.1 & 0.871 & $8.3^{+10.7}_{-8.2}$ $\pm$ 1.5\\ 
1.29(1.20-1.32) &  8.53 & 7.19 & $0.0^{+3.1}_{-1.8}$ & 0.0 $\pm$ 0.1 & 10.0 $\pm$ 2.1 & 0.863 & $0.0^{+5.8}_{-3.4}$ $\pm$ 0.8\\ 
1.37(1.32-1.41) &  6.26 & 7.72 & 11.1  $\pm$ 4.9     & 0.4 $\pm$ 0.1 & 5.0 $\pm$ 1.7 & 0.869 & 26.5 $\pm$ 11.6 $\pm$ 2.4\\ 
1.45(1.41-1.49) &  4.00 & 8.39 & 4.7  $\pm$ 3.5      & 0.7 $\pm$ 0.1 & 4.7 $\pm$ 1.7 & 0.898 & 15.7 $\pm$ 11.6 $\pm$ 2.0\\ 
1.52(1.49-1.59) &  6.52 & 8.98 & 8.4  $\pm$ 6.2      & 2.2 $\pm$ 0.3 & 9.8 $\pm$ 3.0 & 0.936 & 15.3 $\pm$ 11.4 $\pm$ 1.6\\ 
1.60(1.59-1.64) &  3.68 & 8.81 & 9.7  $\pm$ 6.9      & 1.0 $\pm$ 0.2 & 8.0 $\pm$ 3.2 & 0.955 & 31.3 $\pm$ 22.2 $\pm$ 5.4\\ 
1.68(1.64-1.70) &  6.20 & 9.82 & 11.8  $\pm$ 8.5     & 3.0 $\pm$ 0.3 & 12.0 $\pm$ 4.0 & 0.960 & 20.2 $\pm$ 14.6 $\pm$ 3.2\\ 
1.73(1.70-1.75) &  3.60 & 10.98 & 7.1  $\pm$ 5.2     & 1.7 $\pm$ 0.2 & 3.5 $\pm$ 2.1 & 0.959 & 18.6 $\pm$ 13.6 $\pm$ 2.5\\ 
1.78(1.75-1.80) &  7.00 & 10.68 & -0.4  $\pm$ 5.9    & 2.2 $\pm$ 0.2 & 9.4 $\pm$ 3.3 & 0.953 & -0.6 $\pm$ 8.3 $\pm$ 1.3\\ 
1.83(1.80-1.84) &  3.99 & 10.90 & $5.1^{+4.5}_{-3.6}$& 1.1 $\pm$ 0.1 & 4.6 $\pm$ 1.8 & 0.954 & $12.2^{+10.7}_{-8.7}$ $\pm$ 1.0\\ 
1.87(1.84-1.88) & 15.64 & 9.31 & 26.6  $\pm$ 8.7     & 3.0 $\pm$ 0.3 & 14.2 $\pm$ 3.8 & 0.951 & 19.2 $\pm$ 6.3 $\pm$ 1.6\\ 
1.89(1.88-1.92) & 11.43 & 9.80 & 10.9  $\pm$ 7.1     & 2.9 $\pm$ 0.2 & 15.9 $\pm$ 3.4 & 0.948 & 10.3 $\pm$ 6.7 $\pm$ 1.1\\ 
1.97(1.92-2.00) & 12.80 & 9.53 & 1.8  $\pm$ 5.1      & 1.5 $\pm$ 0.2 & 10.5 $\pm$ 2.7 & 0.949 & 1.6 $\pm$ 4.4 $\pm$ 0.6\\ 
\hline
\end{tabular}
\end{table*}

\begin{table*}[p]
\center
\caption{\normalsize 
Weighted average energy ($\sqrt{s}$) for the interval indicated in parentheses,
and the total $e^+e^- \to \eta\pi^0\gamma$ cross section
($\sigma_{\eta\pi\gamma}$). For the cross section, the first error is 
statistical, the second is systematic. 
\label{epgtottab}}
\begin{tabular}{|c|c|c|c|}
\hline
$\sqrt{s}$, GeV & $\sigma_{\eta\pi\gamma}$, pb & $\sqrt{s}$, GeV & $\sigma_{\eta\pi\gamma}$, pb  \\  \hline 
1.15(1.05-1.20) & $8.3^{+10.7}_{-8.2}\pm 1.5$  & 1.73(1.70-1.75) & $127 \pm 20 \pm 9 $    \\  
1.29(1.20-1.32) & $0.0^{+5.8}_{-3.4}\pm 0.7$   & 1.78(1.75-1.80) & $56 \pm 11 \pm 4 $     \\  
1.38(1.32-1.41) & $31 \pm 11 \pm 2$            & 1.83(1.80-1.84) & $17.6 \pm 8.7 \pm 1.3$ \\  
1.45(1.41-1.49) & $33 \pm 13 \pm 2$            & 1.87(1.84-1.88) & $19.2 \pm 5.4 \pm 1.4$ \\  
1.53(1.49-1.59) & $75 \pm 15 \pm 5$            & 1.89(1.88-1.92) & $12.5 \pm 5.8 \pm 1.3$ \\  
1.60(1.59-1.64) & $151 \pm 31 \pm 11$          & 1.96(1.92-2.00) & $3.3 \pm 4.1 \pm 0.6$  \\
1.68(1.64-1.70) & $235 \pm 24 \pm 15$          & &\\
\hline
\end{tabular}
\end{table*}

The main sources of the systematic uncertainty are listed in Table~\ref{syser}.
A significant contribution to the systematic uncertainty arises from the
interference of the $e^+e^-\to\omega\eta$ amplitude with $e^+e^-\to\rho\eta$
amplitude. The $\pi^0\gamma$ mass distribution for the interference term 
has a peak in the $\omega$ meson mass region. Therefore, in this analysis,
the interference will increase or decrease the number of selected events of 
the process $e^+e^-\to\omega\eta$.
The absolute value of the amplitude of the process $e^+e^-\to\rho\eta$ is 
extracted from the measured $e^+e^-\to\pi^+\pi^-\pi^0$ cross 
section~\cite{BaBar_rhoeta}, but the relative 
phase between the two amplitudes is unknown. We calculated 
the total cross section for the processes $e^+e^-\to\omega\eta$ and 
$e^+e^-\to\rho\eta$ with and without interference. The maximum difference
between the two cross sections when varying the phase is taken as an estimate
of the systematic uncertainty of the $e^+e^-\to\omega\eta$ cross section.
In the energy range under study it varies from 5 to 37\%.
The other sources of the systematic uncertainty are discussed in
the previous sections.
\begin{figure}
\centering
\resizebox{0.48\textwidth}{!}{\includegraphics{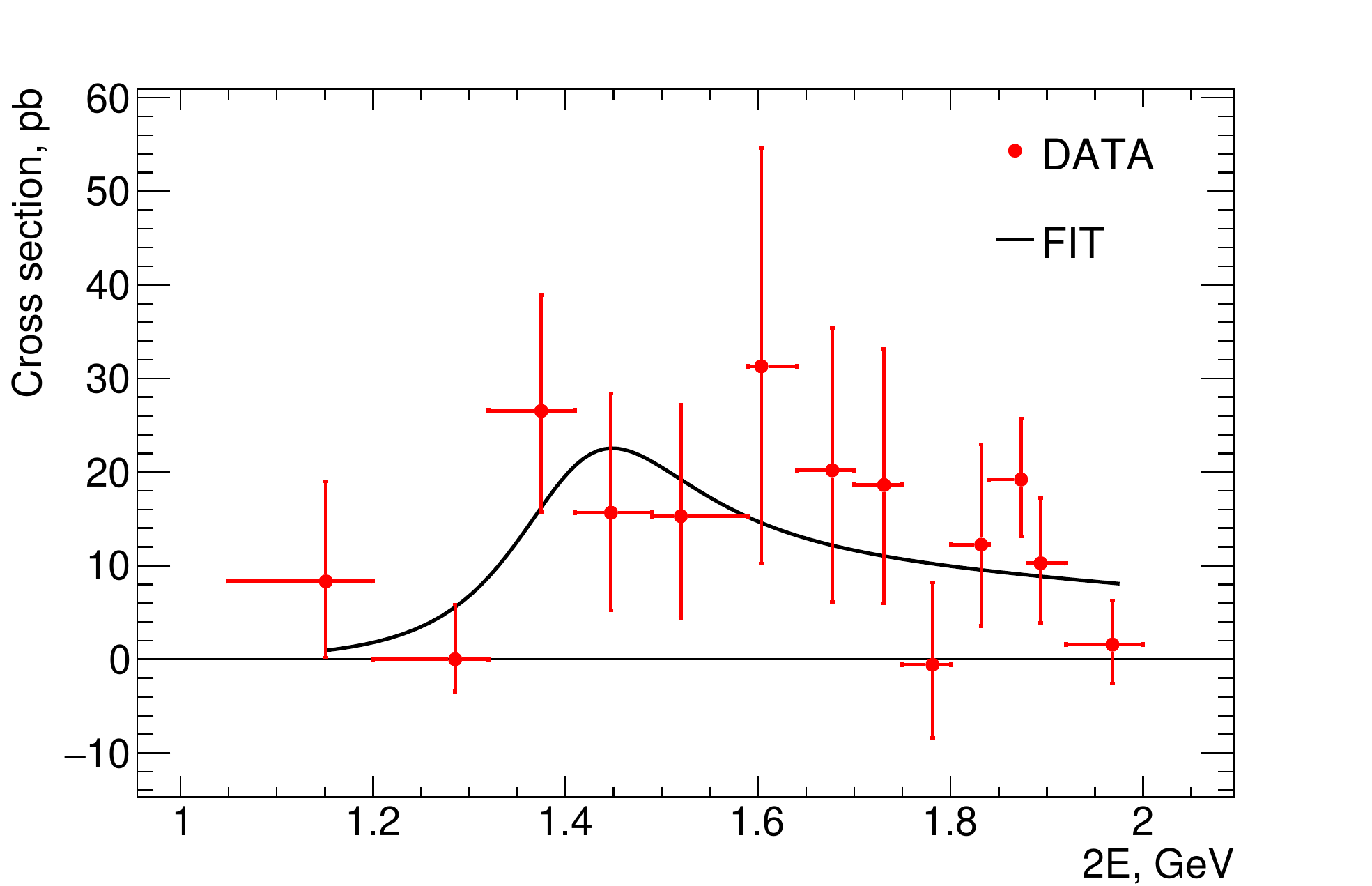}}\hfill
\caption{The measured Born cross section for the process
$e^+e^-\to {\rm rad}\mbox{-}\eta\pi^{0}\gamma$ (points with error bars).
The curve is the result of the fit described in the text.
\label{NonOmeCrsPic}}
\end{figure}

To describe the cross section for the class $\rm rad$-$\eta\pi\gamma$,
we use a model with one vector resonance decaying into $a_0(1420)\gamma$.
The energy dependence of the phase space is calculated using a Monte-Carlo 
event generator for the process $e^+e^-\to a_0(1420)\gamma$. The free
parameters of the fit to the $e^+e^-\to {\rm
rad}\mbox{-}\eta\pi^{0}\gamma$ visible cross section are
$m_{V^{\prime}}$, $\Gamma_{V^{\prime}}$, $B_{V^{\prime}}$. Their fitted
values are listed in Table~\ref{CrsFitRes}.  The values of the radiative 
corrections determined from the fit and the values of the Born cross section
for the process $e^+e^-\to{\rm rad}\mbox{-}\eta\pi\gamma$
with statistical and systematic errors are listed in 
Table~\ref{nonomegatab}. The main sources of systematic uncertainty 
are listed in Table~\ref{syser}. The fitted Born cross
section is shown in Fig.~\ref{NonOmeCrsPic}. It is seen that events of the 
radiative processes are distributed over a wide energy range, 
from 1.3 to 1.9 GeV.
\begin{figure}
\centering
\resizebox{0.48\textwidth}{!}{\includegraphics{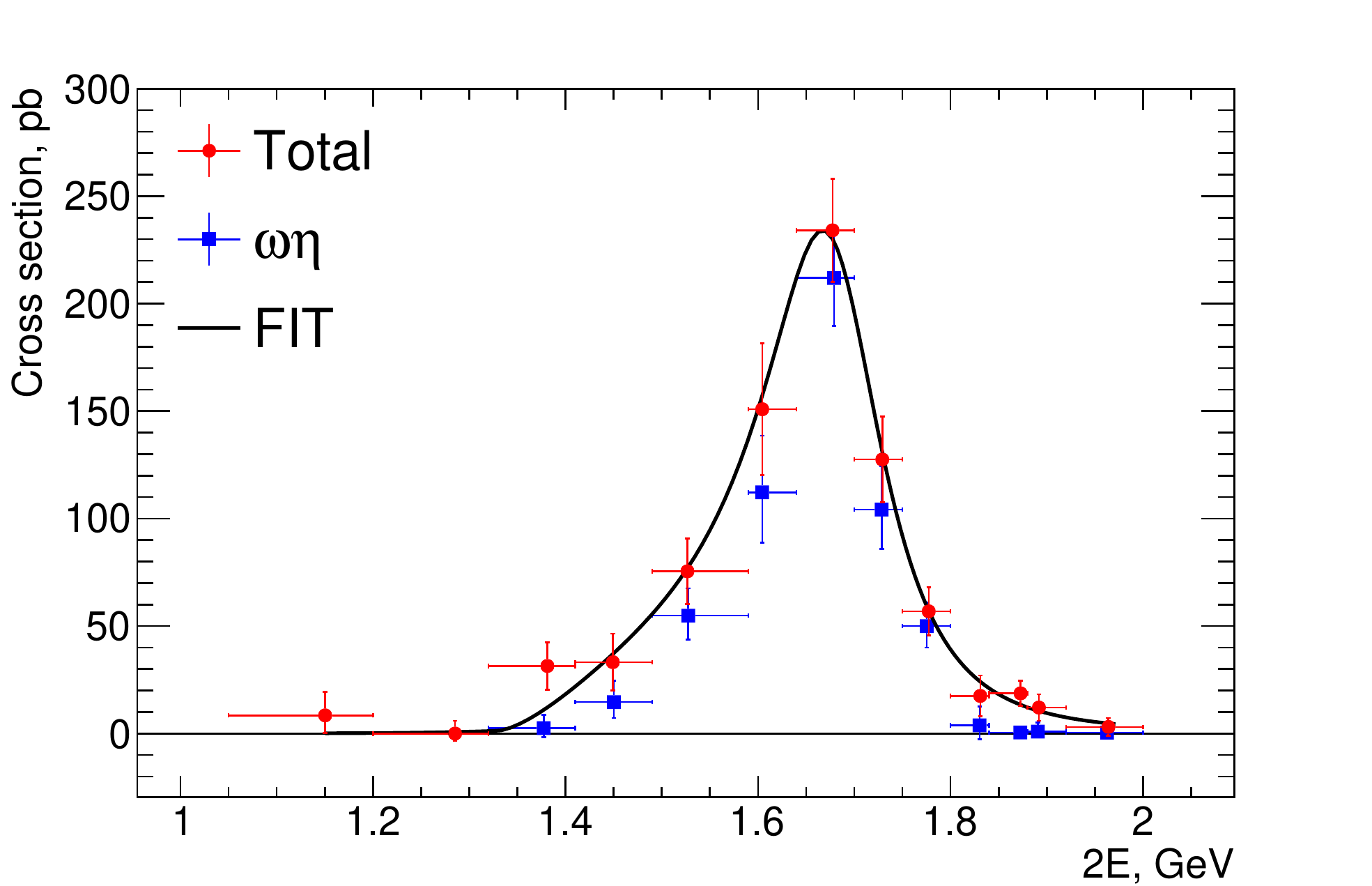}}\hfill
\caption{The measured total cross section for the process $e^+e^-\to\eta\pi^{0}\gamma$ 
(circle) and the cross section for the process 
$e^+e^-\to\omega\eta\to\eta\pi^{0}\gamma$
(squares). The curve is the result of the fit described in the text.
\label{TotCrsPic}}
\end{figure}

The total visible cross section for the process $e^+e^-\to\eta\pi^0\gamma$
is calculated as
\begin{equation}
\sigma_{\rm vis}=\frac{1}{L}
\left(N_{\omega\eta}+
\frac{\varepsilon_{\omega\eta}}{\varepsilon_{\rm rad}}N_{\rm rad}+
\frac{\varepsilon_{\omega\eta}}{\varepsilon_{\rm res}}N_{\rm res}
\right)
\end{equation}
and then fitted by Eq.~(\ref{viscrs}) with the efficiency obtained for
the process $e^+e^-\to\omega\eta$. The model for the Born cross section used
in the fit is a sum of the models for the $e^+e^-\to\omega\eta$ 
and $e^+e^-\to{\rm rad}\mbox{-}\eta\pi\gamma$ cross section. The obtained
values of the Born cross section are listed in Table~\ref{epgtottab} and shown
in Fig.~\ref {TotCrsPic} in comparison with the cross section for the
intermediate state $\omega\eta$. The main sources of the systematic uncertainty
on the cross section are listed in Table~\ref{syser}.
The total cross section for the process $e^+e^-\to\eta\pi^{0}\gamma$ measured
by the described method includes the contribution from the interference of 
the $e^+e^-\to\omega\eta$ and $e^+e^-\to\rho\eta$ amplitudes. Therefore,
the model uncertainty associated with the interference is absent.

\section{Summary}
In the experiment with the SND detector at the VEPP-2000 collider the cross
section for the process $e^+e^-\to\eta\pi^0\gamma$ has been  measured for the
first time in the energy range from 1.05 to 2.00 GeV. The main contribution to the 
cross section arises from the intermediate mechanism $\omega\eta$. The 
measured cross section for the subprocess 
$e^+e^-\to\omega\eta\to\eta\pi^0\gamma$ agrees well with previous measurements
of this cross section by SND and CMD-3 in the decay mode 
$\omega\to\pi^+\pi^-\pi^0$. The significantly smaller contribution to the
$e^+e^-\to\eta\pi^0\gamma$ cross section from other hadronic intermediate 
states $\rho\eta$, $\phi\eta$, $\phi\pi^0$, $\omega\pi^0$, and $\rho\pi^0$ has
been calculated using existing data on their production cross section.
It has been found, with a significance of 5.6$\sigma$, that the process 
$e^+e^-\to\eta\pi^0\gamma$ is not completely described by the hadronic
intermediate states. We assume that the missing contribution
($\rm rad$-$\eta\pi\gamma$) arises from radiation processes, for example,
$e^+e^-\to a_{0}(980)\gamma$, $a_{0}(1450)\gamma$, and $a_{2}(1320)\gamma$.
The cross section for the process $e^+e^-\to{\rm rad}\mbox{-}\eta\pi^0\gamma$ 
has been measured. It is 15--20 pb in a wide energy range, from 1.3 to 1.9 GeV.
The spectrum of $\eta\pi^0$ invariant masses for the events 
$\rm rad$-$\eta\pi\gamma$ is consistent with the dominance of the 
intermediate mechanism $a_{0}(1450)\gamma$.

Our result on the $e^+e^-\to{\rm rad}\mbox{-}\eta\pi^0\gamma$ cross section
can be compared with the predictions of Ref.~\cite{kalashnikova},
where the partial widths for the decays $\rho(1450)$ and  $\omega(1420)$ to
$a_2(1320)\gamma$, and $\omega(1420)$ and $\omega(1650)$ to $a_2(1320)\gamma$
and $a_0(1420)\gamma$ are calculated in the framework of the quark model.
Using the data on the $e^+e^-\to\pi^+\pi^-\pi^+\pi^-$~\cite{4pic_Babar},
$e^+e^-\to\pi^+\pi^-\pi^0\pi^0$)~\cite{4pic_Babar} we estimate that
the total production cross sections for $\rho(1450)$ and $\rho(1700)$ are
about 60 nb and 15 nb, respectively. The production cross section for the
isoscalar resonances $\omega(1420)$ and $\omega(1650)$ are estimated to be
about 6 nb and 9 nb, respectively, from the data on the
$e^+e^-\to\pi^+\pi^-\pi^0$~\cite{3pi_SND},
$e^+e^-\to\omega\pi^+\pi^-$~\cite{om2pic_Babar}, and 
$e^+e^-\to\omega\pi^0\pi^0$~\cite{om2pi_Babar} cross sections. Taking into 
account the branching fractions $B(a_0\to\eta\pi)=0.093\pm 0.020$ and 
$B(a_2\to\eta\pi)=0.145\pm 012$~\cite{pdg}, 
we obtain the cross sections listed in Table~\ref{crs_estimation}. 
Since resonances involved into these subprocesses listed in
Table~\ref{crs_estimation} are wide, we expect
a significant interference between their amplitudes. So, the total
$e^+e^-\to{\rm rad}\mbox{-}\eta\pi^0\gamma$ cross section based on the 
prediction from Ref.~\cite{kalashnikova} may reach 3--5 pb. This value
is several times lower than our measurement.

\begin{table*}[p]
\center
\caption{The predicted contributions to the $e^+e^-\to\eta\pi^0\gamma$ cross
section from the radiative decays of the exited vector mesons in pb. 
The predictions are based on the partial widths calculated in
Ref.~\cite{kalashnikova}.\label{crs_estimation}}
\begin{tabular}{|c|c|c|c|c|}
\hline
& $\rho(1450)$ & $\omega(1420)$ & $\rho(1700)$  & $\omega(1650)$ \\  \hline 
$a_0(1450)\gamma$ & --- & --- & 0.5 & 1.6 \\
$a_2(1320)\gamma$ & 1.3 & 1.2 & 0.1 & 0.4 \\
\hline
\end{tabular}
\end{table*}

\section{Acknowledgments}
The authors are grateful to A.S.Rudenko for useful discussions.

\end{document}